\documentclass[atoms,article,accept,10pt]{mdpi}

\firstpage{1}
\pubvolume{7}
\issuenum{1}
\articlenumber{54}
\pubyear{2019}
\copyrightyear{2019}
\history{Received: 1 May 2019; Accepted: 26 May 2019; Published: 4 June 2019}

\usepackage{longtable}
\usepackage{bm}% bold math
\usepackage{upgreek}

\Title{Transition Rates for $3s3p^2\;^4P - 3s3p4s\;^4P^o$ 
  Transitions in Al {\sc i}
 }

\Author{Charlotte Froese Fischer %
 $^{1}$\orcidA{} and James F. Babb $^2$\orcidB{}}

\address{%
$^{1}$ \quad Department of Computer Science, University of British Colombia, 2366 Main Mall,
Vancouver, BC V6T1Z4, Canada; cff@cs.ubc.ca\\
$^{2}$ \quad Institute for Theoretical Atomic, Molecular, and Optical Physics (ITAMP), Center for Astrophysics|Harvard \& Smithsonian, 60 Garden St., MS 14, Cambridge, MA 02138, USA; jbabb@cfa.harvard.edu}

\corres{Correspondence: cff@cs.ubc.ca; Tel.: +1-604-225-5147. jbabb@cfa.harvard.edu; Tel.: +1-617-496-7612}

\abstract{Fully relativistic calculations have been performed for two multiplets, $3s3p^2\;^4P$ and $3s3p4s\;^4P^o$, in Al {\sc i}. Wave functions were obtained for all levels of these multiplets using the {\sc grasp} 
programs. Reported are the E1 transitions rates for all transitions between levels of these multiplets. Transition energies and transition rates are compared with observed values and other theory. Our~calculated transition rates {are smaller by about 10\% than} observed rates, reducing a large discrepancy between
earlier calculations and experiments.
}

 %%% Please add.

\begin{document}

%%%%%%%%%%%%%%%%%%%%%%%%%%%%%%%%%%%%%%%%%%

\section{Introduction}
{Atomic spectra are of vital importance as plasma diagnostics and reliable
wavelengths and transition probabilities are essential for applications.}
Recently, Hermann~et~al.~\cite{Hermann} deduced transition probabilities
between the fine structure lines connecting the $3s3p^2\;^4P$ and $3s3p^2\;^4P^o$ multiplets from emission coefficients
measured in laser ablation of aluminum in argon.
The transition probabilities obtained were roughly a factor of two larger than those listed in
the semi-empirical calculations of Kurucz and Peytremann~\cite{Kurucz}, and no other values
were found in existing tabulations.
The fine structure lines connecting the $3s3p^2\;^4P$ and $3s3p^2\;^4P^o$ multiplets occur in the same general
wavelength region (305--310~nm) and with comparable strength as the well-studied (see~\cite{ref-Papoulia} and the references therein) fine structure lines connecting
the $3s^23p\;^2P$ and $3s^23d\;^2D$ multiplets, making the discrepancy of {concern} for applications in the UV~wavelengths.

Multiconfiguration Dirac--Hartree--Fock (MCDHF) and relativistic configuration interaction (RCI) calculations have been performed by Papoulia~et~al.~\cite{ref-Papoulia} for 28 states in neutral Al. The~configurations of interest were $3s^2nl$ for $n= 3, 4,5$ with $l$ = 0--4, as~well as $3s3p^2$ and $3s^26l$ for $l=0, 1, 2$. Lifetimes~and transition data for radiative electric dipole (E1) transitions were reported.
There was a significant improvement in accuracy, in~particular for the more complex system of neutral Al {\sc i}, 
which may prove
useful for astrophysical applications to Al abundance determinations in stars.
Omitted were the levels of the $3s3p4s\;^4P^o$ multiplet, which lies above the first $3s^2$ ionization limit~\cite{ErikssonIsberg1963}.

This paper reports transition rates for all E1 transitions between the $3s3p^2\;^4P$ and $3s3p4s\;^4P^o$ multiplets using the variational multiconfiguration Dirac--Hartree--Fock (MCDHF) method~\cite{review}, as~implemented in the {\sc grasp} 
programs~\cite{GRASP}. The~accuracy of the results is based on the accuracy of the theoretically-predicted transition energies {compared with available measurements,} as well as the agreement between length and velocity~rates.

%%%%%%%%%%%%%%%%%%%%%%%%%%%%%%%%%%%%%%%%%%
\section{Underlying~Theory}

In the MCDHF method~\cite{review}, the~wave function $\Psi({ \gamma} PJM_J)$ for a
state labeled ${ \gamma} PJM_J$, where $J$ and $M_J$ are the angular quantum numbers and $P$ the
parity, is expanded in antisymmetrized and coupled configuration state functions (CSFs):
\begin{equation}
\label{ASF}
\Psi({ \gamma} PJM_J) = \sum_{j=1}^{M} c_{j} \Phi(\gamma_{j}PJM_J).
\end{equation}

The labels $\{\gamma_j\}$ denote other appropriate information about the CSFs,
such as orbital occupancy and the coupling scheme.
The CSFs are built from products of one-electron orbitals, having the general~form:
\begin{align}
\label{g2}
\psi_{n\kappa, m}(\mathbf{r}) = \frac{1}{r}
\begin{pmatrix} P_{n\kappa}(r) \chi_{\kappa, m}(\theta,\varphi) \\
\imath Q_{n\kappa}(r) \chi_{-\kappa, m}(\theta,\varphi)
\end{pmatrix},
\end{align}
where $\chi_{\pm \kappa, m}(\theta,\varphi)$ are two-component spin-angular functions. The~expansion coefficients and the radial functions are determined iteratively. In~the present work, the~Dirac--Coulomb Hamiltonian ${\cal H}_{DC}$ was used~\cite{review}, which included a correction for the finite size of the~nucleus.

The radial functions $\{P_{n\kappa}(r), Q_{n\kappa}(r)\}$ were determined numerically as solutions of differential~equations,
\begin{eqnarray}
 \label{eq:DHF-operator}
 \hspace{-2cm}
 w_a
 \left[ \begin{array}{c c}
 V(a;r) & -c\left[ \frac{d}{dr} - \frac{\kappa_a}{r} \right] \\
   \ \\
   c\left[ \frac{d}{dr} + \frac{\kappa_a}{r} \right] & V(a;r) -2c^2
   \end{array}\right]
  \left[\begin{array}{c}
   P_a(r)\\
   Q_a(r)
   \end{array}\right]
   = \sum_{b} \epsilon_{ab} \; \delta_{\kappa_a \kappa_b}
 \left[\begin{array}{c}
   P_b(r)\\
   Q_b(r)
   \end{array}\right],
 \end{eqnarray}
 where $V(a;r) = V_{nuc}(r) + Y(a;r) + \bar{X}(a;r)$ is a potential consisting of nuclear, direct, and~exchange contributions arising from both diagonal and off-diagonal
 $\langle \Phi_\alpha \vert {\cal H}_{DC} \vert \Phi_\beta \rangle$ matrix elements~\cite{review}.

For a given set of radial functions, expansion coefficients ${\bf c} = (c_1,\ldots,c_M)^t$ were obtained as solutions to the configuration interaction (CI) problem,
\begin{equation}
{\bf H}{\bf c} = E {\bf c},
\end{equation}
where ${\bf H}$ is the CI matrix of dimension $M \times M$ with elements
\begin{equation}
H_{ij} = \langle \Phi(\gamma_{i}PJM_J)|H| \Phi(\gamma_{j}PJM_J) \rangle.
\end{equation}

Once self-consistent solutions have been obtained---sometimes referred to as the relativistic MCDHF or RMCDHF 
 phase---an  
 RCI calculation was performed using an extended Hamiltonian that included the transverse photon (Breit) and QED corrections. Wave functions from the latter Hamiltonian were used to compute the E1 transitions~rates.

 \section{Systematic~Procedures}

 Systematic procedures were used in which the orbital set used for defining the wave function expansion increased systematically within a correlation~model.

 The states of Al consist of a neon-like ($1s^22s^22p^6$) core and three valence electrons. Wave~function expansions were obtained from single- and double- (SD) excitations from a multireference (MR) set that interacted significantly with the CSFs of interest. For~the even multiplet, the~MR set included the CSFs from $3s3p^2$, $3s3d^2$, and~$3p^23d$ configurations and for the odd multiplet, CSFs from $3s3p4s$, $3p3d4s$, $3p^24p$, and~$3s3d4p$. Because~the odd multiplet was above the first ionization limit, all $3s^2nl$ CSFs were removed. In~this paper, the~orbital sets that define the set of excitations were classified according to the largest $nl$ of the orbital set when the latter were ordered globally by $n$ (the principal quantum number) and within $n$ by $l$ (the orbital quantum number). Thus, an $n=3$ orbital set includes all orbitals up to $3s, 3p, 3d$ and all $n=5f$ orbitals up to $5s,5p,5d,5f$ ($5g$ {not included}).

Our first model was the valence correlation (VV) model, in which all excitations involved only valence electrons. $n=3$ calculations were performed for an average energy functional of the lowest even parity $J=1/2, 3/2, 5/2$ states. This calculation defined the core orbitals for all subsequent calculations. The~$n=4,5f, 6f$ calculations each varied only the new orbitals. For~the odd multiplet, the~first calculation had orbitals up to $4s$, $4p$, and~$3d$, but is still referred to as an $n=3$ calculation in this paper, with~remaining sets being the regular $n= 4,5f,6f$ orbital sets. Table~\ref{tab:vv} shows the convergence of the fine-structure of the two multiplets and their~separation.

\begin{table}[H]
\centering
\caption{Convergence of the energy level structure for a valence correlation calculation is compared with observed {data}~\cite{ErikssonIsberg1963,NIST}. All results are in cm$^{-1}$.
\label{tab:vv}}
\begin{tabular}{lrrrrr}
\toprule
               &  \boldmath$n=3$ & \boldmath$n=4$  & \boldmath$n=5f$ &  \boldmath$n=6f$ & \textbf{Observed}\\
\midrule
\multicolumn{6}{l}{ $3s3p^2\;^4P$ Fine structure}\\
 $3s3p^2~^4P_{ 1/2 }$~&    0 &    0 &    0 &    0 & 0.00 \\
 $3s3p^2~^4P_{ 3/2 }$~&   46 &   46 &   45 &   45 & 46.55 \\
 $3s3p^2~^4P_{ 5/2 }$~&   122 &   121 &   120 &   120 & 122.37 \\
\hline
\multicolumn{6}{l}{ $3s3p4s\;^4P^o$ Fine structure}\\
$3s3p4s~^4P_{ 1/2 }^o$~&    0 &    0 &    0 &    0 & 0.00 \\
$3s3p4s~^4P_{ 3/2 }^o$~&   57 &   58 &   56 &   55 & 56.10 \\
$3s3p4s~^4P_{ 5/2 }^o$~&   156 &   157 &   153 &   150 & 152.08 \\
\hline
\multicolumn{6}{l} {Multiplet separation}\\
$3s3p4s~^4P_{ 1/2 }^o$ - $3s3p^2~^4P_{ 1/2 }$~&
  33,213 & 33,029 & 32,724 & 32,696 & 32,671.05 \\
\bottomrule
\end{tabular}
\end{table}

The results of the converged valence correlation calculations, when compared with observation, suggest that the energy structure is not significantly affected by the core-valence (CV) that accounts for the polarization for the core. To~confirm this conclusion, calculations were performed in which SD excitations included a single excitation from the $2p$-shell along with a single excitation of a valence electron. Wave function expansions were considerably larger and convergence a bit slower. Table~\ref{tab:cv} shows the convergence of the energy structure. The~fine structure of the odd multiplet increased slightly and was in somewhat better agreement with the data from observation. At~the same time, the~transition energy for $^4P_{1/2} - ^4P^o_{1/2}$ for an $n=7f$
calculation was not in as good agreement with observed as the $n=6f$ valence
correlation calculation reported in Table~\ref{tab:vv}.

\begin{table}[H]
\centering
\caption{Convergence of the energy level structure from a core-valence plus valence correlation calculation is compared with observed data~\cite{ErikssonIsberg1963,NIST}. All results are in cm$^{-1}$.
\label{tab:cv}}
\begin{tabular}{lrrrrr}

\toprule
     & \boldmath$n=5f$  & \boldmath$n=6f$ &  \boldmath$n=7f$ & \textbf{Observed}\\
\midrule
\multicolumn{5}{l}{ $3s3p^2\;^P$ Fine structure}\\
$3s3p^2~^4P_{ 1/2 }$~&    0 &    0 &    0 & 0.00 \\
$3s3p^2~^4P_{ 3/2 }$~&   48 &   45 &   45 & 46.55 \\
$3s3p^2~^4P_{ 5/2 }$~&   128 &   120 &   120 & 122.37 \\\hline
\multicolumn{5}{l}{$3s3p4s\;^4P^o$ Fine structure}\\
$3s3p4s~^4P_{ 1/2 }^o$~&    0 &    0 &    0 & 0.00 \\
$3s3p4s~^4P_{ 3/2 }^o$~&   60 &   56 &   56 & 56.10 \\
$3s3p4s~^4P_{ 5/2 }^o$~&   167 &   155 &   153 &152.08 \\
\hline
\multicolumn{5}{l} {Multiplet separation}\\
$3s3p^2~^4P_{ 1/2 }$-$3s3p4s~^4P_{ 1/2 }^o$ & 33,142.58 & 32,934.21 & 32,838.33 & 32,671.05 \\
\bottomrule
\end{tabular}
\end{table}

%%%%%%%%%%%%%%%%%%%%%%%%%%%%%%%%%%%%%%%%%%
\section{Results}

The wave functions from RCI expansions, determined using the Dirac--Coulomb--Breit-QED Hamiltonian, were used to compute the E1 transition rates for all transitions between these two multiplets. Table~\ref{tab-trvv} reports the transition energy $\Delta E$ (cm$^{-1}$), the~wavelength $\lambda$ (nm) in a vacuum, $A$~($\upmu$s$^{-1}$), and $gf$, in~the length form for calculations of Table~\ref{tab:vv}. Furthermore included is an indicator of accuracy $dT = (A_l - A_v)/\mathrm{max}(A_l, A_v)$, where $A_l$ and $A_v$ are transition rates from length and velocity forms, respectively. The~average discrepancy between the two forms was 1.5\%, and in~all cases, the~velocity form had a larger value than the length~form.

\begin{table}[H]
\centering
\caption{Ab~initio electric dipole (E1) transition
data for the $3s3p^2\;^4P$ to $3s3p4s\;^4P^o$ transition computed in
the length form\label{tab-trvv} {from valence correlation results.}}
\begin{tabular}{lllrrrr} 
\toprule
\textbf{Upper} & \textbf{Lower} & 
 \boldmath$\Delta E$ \textbf{(cm$^{-1}$)} & \textbf{\boldmath$\lambda$ (nm)} &
\boldmath$A$ \textbf{($\upmu$s$^{-1}$)} & \boldmath$gf$ & \boldmath$dT$               \\ \midrule
$3s3p4s~^4P_{ 1/2 }^o$ & $3s3p^2~^4P_{ 1/2 }$ &  32,696 & 305.84220 & 29.93 & 0.0840 &   0.016\\
$3s3p4s~^4P_{ 1/2 }^o$ & $3s3p^2~^4P_{ 3/2 }$ &  32,650 & 306.27027 & 149.01 & 0.4191 &   0.017\\
$3s3p4s~^4P_{ 3/2 }^o$ & $3s3p^2~^4P_{ 1/2 }$ &  32,752 & 305.32421 & 75.11 & 0.4199 &   0.014\\
$3s3p4s~^4P_{ 3/2 }^o$ & $3s3p^2~^4P_{ 3/2 }$ &  32,706 & 305.75074 & 23.89 & 0.1339 &   0.015\\
$3s3p4s~^4P_{ 3/2 }^o$ & $3s3p^2~^4P_{ 5/2 }$ &  32,631 & 306.44982 & 80.43 & 0.4529 &   0.017\\
$3s3p4s~^4P_{ 5/2 }^o$ & $3s3p^2~^4P_{ 3/2 }$ &  32,801 & 304.86634 & 54.37 & 0.4546 &   0.014\\
$3s3p4s~^4P_{ 5/2 }^o$ & $3s3p^2~^4P_{ 5/2 }$ &  32,726 & 305.56128 & 125.94 & 1.0577 &   0.015\\ \bottomrule
\end{tabular}
\end{table}

The effect of core-valence is shown in Table~\ref{tab-trcv} where results are reported both for the $n=5f$ and $n=7f$ calculation. The~former was included because of the remarkable agreement in length and velocity forms, yet the transition energy (as shown in Table~\ref{tab:cv}) was not in as good agreement with the observed as before. Since the transition rate is proportional to $(\Delta E)^3$, correcting the transition rate for this factor would introduce a 4.0\% reduction. Indeed, the~$n=7f$ transition rates were smaller with the transition energy more accurate, but length and velocity were not in as good~agreement.

\begin{table}[H] 
\centering
\caption{Ab~initio electric dipole (E1) transition data for the $3s3p^2\;^4P$ to $3s3p4s\;^4P^o$ transition computed in
the length form\label{tab-trcv} {from valence and core-valence results}.}
\begin{tabular}{llrrrrr} 
\toprule
\textbf{Upper} & \textbf{Lower }& \boldmath
$\Delta E$ \textbf{(cm$^{-1}$) }& \boldmath$\lambda$ \textbf{(nm)} &
\boldmath$A$ \textbf{($\upmu$s$^{-1}$) }& \boldmath$gf$ & \boldmath$dT$               \\ \midrule

\multicolumn{7}{l}{ $n=5f$ }\\\hline
$3s3p4s~^4P_{ 1/2 }^o$ & $3s3p^2~^4P_{ 1/2 }$ &  33,142 & 301.727 & 29.76 & 0.0812 & 0.004\\
$3s3p4s~^4P_{ 1/2 }^o$ & $3s3p^2~^4P_{ 3/2 }$ &  33,093 & 302.170 & 148.19 & 0.4057 & 0.006\\
$3s3p4s~^4P_{ 3/2 }^o$ & $3s3p^2~^4P_{ 1/2 }$ &  33,203 & 301.173 & 74.72 & 0.4064 & 0.001\\
$3s3p4s~^4P_{ 3/2 }^o$ & $3s3p^2~^4P_{ 3/2 }$ &  33,154 & 301.615 & 23.88 & 0.1303 & 0.003\\
$3s3p4s~^4P_{ 3/2 }^o$ & $3s3p^2~^4P_{ 5/2 }$ &  33,075 & 302.341 & 79.91 & 0.4380 & 0.008\\
$3s3p4s~^4P_{ 5/2 }^o$ & $3s3p^2~^4P_{ 3/2 }$ &  33,261 & 300.652 & 53.83 & 0.4377 & 0.001\\
$3s3p4s~^4P_{ 5/2 }^o$ & $3s3p^2~^4P_{ 5/2 }$ &  33,181 & 301.373 & 125.12 & 1.0223 & 0.003\\\hline
\multicolumn{7}{l}{ $n=7f$ }\\\hline
$3s3p4s~^4P_{ 1/2 }^o$ & $3s3p^2~^4P_{ 1/2 }$ &  32,838 & 304.522 & 28.51 & 0.0793 & 0.051\\
$3s3p4s~^4P_{ 1/2 }^o$ & $3s3p^2~^4P_{ 3/2 }$ &  32,792 & 304.947 & 141.94 & 0.3958 & 0.053\\
$3s3p4s~^4P_{ 3/2 }^o$ & $3s3p^2~^4P_{ 1/2 }$ &  32,894 & 304.003 & 71.57 & 0.3966 & 0.049\\
$3s3p4s~^4P_{ 3/2 }^o$ & $3s3p^2~^4P_{ 3/2 }$ &  32,848 & 304.426 & 22.85 & 0.1270 & 0.050\\
$3s3p4s~^4P_{ 3/2 }^o$ & $3s3p^2~^4P_{ 5/2 }$ &  32,773 & 305.120 & 76.55 & 0.4274 & 0.055\\
$3s3p4s~^4P_{ 5/2 }^o$ & $3s3p^2~^4P_{ 3/2 }$ &  32,946 & 303.523 & 51.61 & 0.4277 & 0.047\\
$3s3p4s~^4P_{ 5/2 }^o$ & $3s3p^2~^4P_{ 5/2 }$ &  32,871 & 304.213 & 119.86 & 0.9979 & 0.050\\
\bottomrule
\end{tabular}
\end{table}

In summary, valence correlation predicted the best transition energy and the best agreement in length and velocity for accurate transition~energy.

{\section{Summary and~Conclusions}}

Table~\ref{compare} compares the predicted transition rates (based on observed transition energies or wavelengths in a vacuum, {rather than computed transition energies as in Table~\ref{tab-trcv}}), with values derived from observations by Hermann~et~al.~\cite{Hermann} and the values reported by Kurucz and Peytremann~\cite{Kurucz}. The~latter used
a semi-empirical approach
in which Slater parameters were determined empirically from observed energy levels and transition probabilities calculated
by the use of scaled Thomas--Fermi--Dirac wave functions. As~seen in Table~\ref{compare}, the~present predicted transition rates are about 10\% smaller than observed values, whereas the Kurucz and Peytremann values are about a half those of the observed rates. Thus, the discrepancy between theory and experiment has been reduced~significantly.

\begin{table}[H]
\centering
\caption{ Comparison of the transition rates {computed} from valence correlation calculations (present) and observed wavelengths (in a vacuum)
from NIST~\cite{ErikssonIsberg1963,NIST} with observed rates
from Hermann~et~al.~\cite{Hermann} and values reported by Kurucz and Peytremann~\cite{Kurucz} 
\label{compare} }
\begin{tabular}{@{\extracolsep{4pt}}lllrrrrr@{}}
\toprule
\textbf{Upper} & \textbf{Lower} & \multicolumn{2}{c}{\boldmath$\lambda$\textbf{ (nm)}} &
\multicolumn{3}{c}{\boldmath$A$ ($\upmu$s$^{-1}$)} \\
\cline{3-4}\cline{5-7}
  & & \textbf{NIST} & \textbf{Present} & \textbf{Present} & \textbf{Hermann} & \textbf{Kurucz} \\ \midrule
$3s3p4s~^4P_{ 1/2 }^o$ & $3s3p^2~^4P_{ 1/2 }$ & 305.9924 & 305.8422 & 29.89 & 34. & 18.3\\
$3s3p4s~^4P_{ 1/2 }^o$ & $3s3p^2~^4P_{ 3/2 }$ & 306.4290 & 306.2703 & 148.80 & 160. & 89.2\\
$3s3p4s~^4P_{ 3/2 }^o$ & $3s3p^2~^4P_{ 1/2 }$ & 305.4679 & 305.3242 & 74.99 & 78. & 44.9\\
$3s3p4s~^4P_{ 3/2 }^o$ & $3s3p^2~^4P_{ 3/2 }$ & 305.9029 & 305.7507 & 23.85 & 28. & 14.2\\
$3s3p4s~^4P_{ 3/2 }^o$ & $3s3p^2~^4P_{ 5/2 }$ & 306.6144 & 306.4498 & 80.32 & 90. & 47.7\\
$3s3p4s~^4P_{ 5/2 }^o$ & $3s3p^2~^4P_{ 3/2 }$ & 305.0073 & 304.8663 & 54.28 & 59. & 32.1\\
$3s3p4s~^4P_{ 5/2 }^o$ & $3s3p^2~^4P_{ 5/2 }$ & 305.7144 & 305.5613 & 125.75 & 140. & 75.0\\
\midrule
\end{tabular}
\end{table}
%%%

%%%%%%%%%%%%%%%%%%%%%%%%%%%%%%%%%%%%%%%%%%
\authorcontributions{The authors C.F.F. and J.F.B. contributed jointly to conceptualization, methodology, validation, formal analysis,  writing--original draft preparation, and writing--review and editing.}

\funding{This research was funded by Canada NSERC Discovery Grant 2017-03851 (CFF)
and US NSF Grant No.\ PHY-1607396 (JFB). The APC was funded by MDPI.} 

%%%%%%%%%%%%%%%%%%%%%%%%%%%%%%%%%%%%%%%%%%
\acknowledgments{The authors (CFF and JB) acknowledge support,
respectively, from~the Canada NSERC Discovery Grant 2017-03851 and
from ITAMP
, which
is supported in part by Grant No.\ PHY-1607396 from the NSF to Harvard University and the Smithsonian Astrophysical Observatory.}

%%%%%%%%%%%%%%%%%%%%%%%%%%%%%%%%%%%%%%%%%%

\conflictsofinterest{The authors declare no conflict of interest. The funders had no role in the design of the study; in the collection, analyses, or interpretation of data; in the writing of the manuscript, or in the decision to publish the results.} 

\reftitle{References}


\begin{thebibliography}{999}

\bibitem{Hermann}
Hermann, J.; Lorusso, A.; Perrone, A.; Strafella, F.; Dutouquet, C.; Torralba, B. Simulation of emission spectra from nonuniform reactive laser-induced plasmas. {\em Phys. Rev. E} \textbf{2015},
{\emph{92}}, 053103.

\bibitem{Kurucz}
Kurucz, R.L.; Peytremann, E. {\em Research in Space Science, SAO Special Report No. 362}; Smithsonian Institution: Astrophysical Observatory: Cambridge, MA, USA, 1975.


\bibitem{ref-Papoulia}
Papoulia, A.; Ekman, J.; J\"onsson, P. Extended transition rates and lifetimes in Al I and Al II from systematic multiconfiguration calculations. {\em A\&A } \textbf{2019}, {621}, A16. 


\bibitem{ErikssonIsberg1963}
Eriksson, K.B.S.; Isberg, H.B.S. The spectrum of atomic aluminium, Al I. {\em Ark. Fys. } \textbf{1963}, \emph{23}, 527.


\bibitem{review}
Fischer, C.F.; Godefroid, M.; Brage, T.; J\"onsson, P.; Gaigalas, G. Advanced multiconfiguration methods for complex atoms: I. Energies and wave functions. {\em J. Phys. B At. Mol. Opt. Phys.} {\bf 2016}, \emph{49}, 182004.

\bibitem{GRASP}
Fischer, C.F.; Gaigalas, G.; J\"onsson, P.; {Biero\'{n}}, J. GRASP2018-A Fortran 95 version of the General Relativistic Atomic Structure Package.
 {\em Comput. Phys. Commun.} \textbf{2019}, {\emph{237}}, 184--187.

\bibitem{NIST}
Kramida, A.; { Ralchenko}, Y.; Reader, J.; NIST ASD Team.
\emph{ NIST Atomic Spectra Database
(ver. 5.5.6)}; National Institute of Standards and Technology:
Gaithersburg, MD, USA, 2019. Available online:
\url{https://physics.nist.gov/asd} (accessed on 15 May 2019).




\end{thebibliography}
\end{document}